# Elimination of the confrontation between theory and experiment in flexoelectric $Bi_2GeO_5$

Yuying Cao[1], Xulong Zhang[1], Long Zhou[1], Hongfei Liu[1], Hua Gao[1], Fu Zheng[1], Zhi Ma*[1,2,3]

[1] *School of Physics, Ningxia University, Yinchuan 750021, China*

[2] *State Key Laboratory of High-Efficiency Utilization of Coal and Green Chemical Engineering, Ningxia University, Yinchuan 750021, China*

[3] *Ningxia Key Laboratory of Intelligent Sensing for Desert Information, Ningxia University, Yinchuan, 750021, China*

*Corresponding author: mazhi@nxu.edu.cn (Z. Ma);

Tel: 0951-2061430, 13619587231         Fax: 0951-2061430

**Abstract.** In this paper, we have investigated the flexoelectric effect of $Bi_2GeO_5$(BGO), successfully predicted the maximum flexoelectric coefficient of BGO, and tried to explore the difference between experimental and simulated flexoelectric coefficients.

## 1. Introduction

Oxides with high ferroelectric transition temperatures that do not contain the toxic substance lead have attracted interest, and the scientific importance and environmental friendliness of such oxides have motivated their study. In this case, $Bi_2GeO_5$ is a kind of stable and non-toxic lead-free ferroelectric crystal with good ferroelectricity, which has the advantages of high stability, high refractive index, high activity, and special structure. $Bi_2GeO_5$ is ferroelectric and the thermal analysis results observed by Polosan demonstrate the phase transition temperature of 843 K[1]. Brasov also found ferroelectricity at Tc>800 K and the magnitude of the ferroelectricity is comparable to that of barium titanate[2].

Binary $Bi_2O_3$-$GeO_2$, with a $Bi_2O_3$:$GeO_2$ ratio of 1:1 and a crystal structure of $Bi_2GeO_5$, is an excellent optical glass because both $Bi_2O_3$ and $GeO_2$ oxides have large electronic polarizations and optical alkalinity[3,4,5],Therefore high refractive indices are expected in such binary-based glasses. $Bi_2GeO_5$ is a stable n-type semiconductor with a relatively high redox capacity that can be applied to electrolytes, ferroelectrics and oxygen precipitating photocatalysts[6].

The crystal structure of $Bi_2GeO_5$ is monoclinic (close to an orthorhombic crystal system) and can maintain anisotropic crystal shapes in $Bi_2GeO_5$ nanocrystals[1]. As shown in Fig.1, BGO has a crystal structure typical of Aurivillius, consisting of alternating layers of $Bi_2O_2$ ions and $GeO_4$ tetrahedral layers, with the $Bi_2O_2$ layer forming a $BiO_4$ square pyramid. The $GeO_4$ tetrahedral layers, where the four oxygen atoms are located at the vertices of the tetrahedra, and the Ge atom are located at the center of the tetrahedron, are connected in a one-dimensional (1D) chain pattern[7].

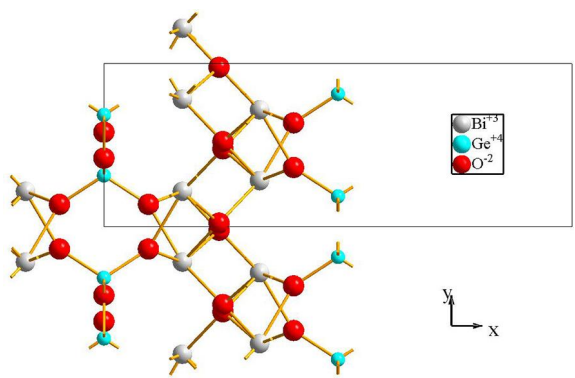

Fig.1 Crystal structure of $Bi_2GeO_5$

In ferroelectric materials, stress can cause a change in the polarization strength, which can be expressed as the flexoelectric effect. The flexoelectric effect is an electromechanical coupling effect that describes the coupling between the strain gradient and the polarization. It differs from the piezoelectric effect in that it is widespread in all insulators in general. The flexoelectric effect becomes important at tiny scales and has attracted much attention, and it has been studied with different research methods. However, a glance at the literature reveals that experimentally measured flexoelectric coefficients are always several orders of magnitude higher than those predicted by simulations. We have chosen the material BGO to investigate its flexoelectric coefficient and analyze the possible reasons for the difference between experiments and simulations.

## 2. Theory and methodology

Computer simulation is an effective problem-solving method. Classical molecular dynamics(MD) is an effective simulation method. Molecular dynamics(MD) works by calculating the molecular potential, which gives the velocity and position of each atom in successive steps. In molecular dynamics, the success of the simulation depends heavily on the choice of the potential function. In this paper, a core-shell model has been used to simulate the effects on ferroelectric materials, which has been widely used in the study of ferroelectric materials[8]. In the shell model, each atom consists of a positively charged ionic shell (with a mass of almost the entire atom) and a negatively charged ionic shell (with almost no mass), and the sum of both charges is the total charge of the atom[8]. Part of the nucleoshell interaction force of the BGO is the shell-shell potential, the Buckingham potential, which describes shell-shell repulsion and van der Waals gravity and can be written as

$$V_{r_{ij}}^{Back} = A\exp(-\frac{|r_{ij}|}{\rho}) - \frac{C}{r_{ij}^6} \tag{1}$$

where $r_{ij}$ is the distance between two nuclei and A, $\rho$, C is the parameter of the potential. The other part is the coulomb potential, which describes the long-range electrostatic interaction between two particles and reflects the core-shell interaction between different particles. The standard Ewald summation procedures for the coulombic interactions are used in the lattice energy calculations. The potential is as following

$$V_{r_{ij}}^{Coulomb} = \frac{q_i q_j}{4\pi\varepsilon_0 |r_{ij}|} \tag{2}$$

where $q_i$ and $q_j$ are the charges of particles i and j respectively, and $\varepsilon_0$ is the vacuum permittivity.

The third part is the interaction between the nucleus-shells of the same atom, it is written as

$$V_{r_{ij}}^{Spring} = \frac{1}{2} k_2 |r_{ij}^2| + \frac{1}{24} k_4 |r_{ij}^4| \tag{3}$$

where $k_2$ and $k_4$ are the potential energy parameters of the non-harmonic spring. In this work, $k_4=0$ and the other parameters of the potential have been listed in Table 1.

Table 1 The parameters for core-shell model of $Bi_2GeO_5$ ferroelectrics[9]

| Short-range interactions | A/eV | $\rho$/nm | C(×10⁻⁶,eV·nm⁶) | Atoms | $k_2$/(×10⁻²,eV·nm²) | shell/e |
|---|---|---|---|---|---|---|
| Bi-O | 49529.35 | 0.2223 | 0 | $Bi^{3+}$ | 359.55 | -5.51 |
| Ge-O | 2090.00 | 0.3172 | 53.7 | $Ge^{4+}$ | 1000 | 4.00 |
| Bi-Bi | 24244.50 | 0.3284 | 0 | $O^{2-}$ | 6.30 | -2.04 |
| O-O | 9547.96 | 0.2192 | 32 | | | |

Molecular dynamics, which allows structural or conformational properties to be extracted from simulated motions of clusters of atoms, is a useful method for studying the polarization of ferroelectrics. The core-shell model potentials for interactions between the ions are widely used in ionic materials. The core-shell potential works well and is computationally efficient, where the charge parameter and the short-range interaction parameter are very important. The radial distribution function gives an idea of the stability of the model and the motion of the molecules. A model with dimensions 6×6×6, with 13824 atoms, size 9.414 nm×3.2952 nm×3.2298 nm was built, in which periodic boundary conditions were used. It uses both periodical boundary conditions and gradients. The NPT ensemble is utilised in all equilibrium processes during the molecular dynamics simulation. The pressure control method utilizes the Nose-Hoover pressure control scheme and the damping parameter is set to 0.4. Unlike the equilibrium process, the radial distribution function and flexoelectric are calculated using the NVE set. For the calculation of flexural electrics, the model size is 2×5×40 and the size is 3.138 nm×2.746 nm×21.532 nm. One section of the cantilever beam is fixed, and the other section is bent with a force of constant magnitude towards y. Periodic boundary conditions are used in the x-direction, and contraction boundary conditions are used in the y- and z-directions.

## 3. Results and discussion

### 3.1 Radial distribution function

It is often common to use g(r) to denote the radial distribution function that describes the motion of the atoms and the distances between them. In order to investigate the microscopic behavior of $Bi_2GeO_5$ at different temperatures as well as to verify its structural stability, supercells with dimensions of 6×6×6 were established and the radial distribution functions of $Bi_2GeO_5$ at five different temperatures, namely, 10 K, 250 K, 500 K, 750 K, and 1000 K, were calculated. The simulation results are shown in Fig. 2.

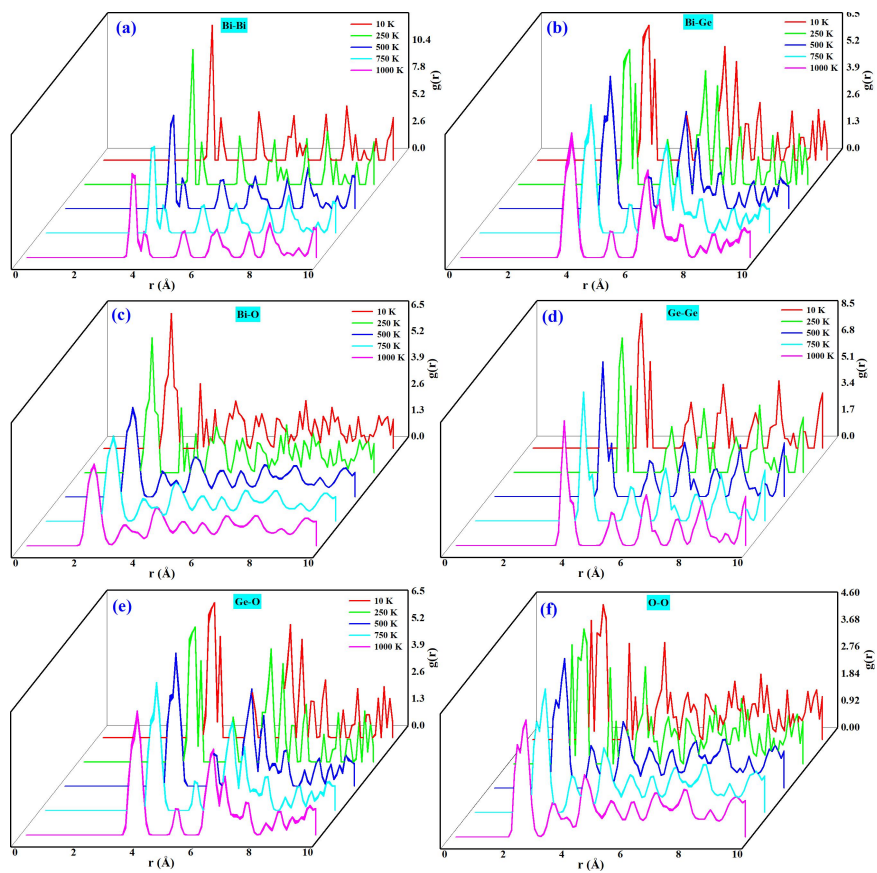

Fig.2 The radial distribution functions for (a)Bi-Bi, (b)Bi-Ge, (c)Bi-O, (d)Ge-Ge, (e)Ge-O, and (f)O-O at temperatures of 10 K, 250 K, 500 K, 750 K, and 1000 K, respectively.

The Fig. 2 indicates that at different temperatures, Bi-Bi, Bi-Ge, Bi-O, Ge-Ge,

Ge-O, and O-O exhibit sharp peaks, suggesting a tightly arranged system of atoms. Upon comparing the peaks of the same paired atoms at different temperatures, it is evident that the value of g(r) is decreasing, and the peak's width is increasing with temperature. This indicates that the lattice vibrations of the atoms become increasingly erratic with temperature. To confirm that the model structure remains stable as temperature increases, the radial distribution function was utilized to calculate the distances between pairs of atoms. By calculating the change in the bond lengths of the atoms at different temperatures, it is found that the bond lengths of the atoms remain constant even when the temperature rises to 1000 K. The bond lengths of Bi-Bi is 3.75 Å, Bi-Ge is 3.85 Å, Bi-O is 2.35 Å, Ge-Ge is 3. 75 Å, Ge-O is 3.85 Å and O-O is 2.45 Å. The bond lengths of atoms at different temperatures are also found to remain constant even when the temperature is increased to 1000 K. It is shown that the model of $Bi_2GeO_5$ remains stable in the temperature range of 10 K-1000 K, paving the way for the next investigations on phase transitions.

## 3.2 Hysteresis loop

Ferroelectric materials are very sensitive to changes in an electric field, and the spontaneous polarization of ferroelectric materials changes in the presence of an applied electric field. To study its anisotropy, a model containing 13,824 atoms with dimensions of 9.414 nm×3.2952 nm×3.2298 nm in size was built. To compare the differences in spontaneous polarization in different directions of [001],[110] and [111], an electric field of up to 2000 MV/m was added in different direction. The hysteresis loop of BGO was shown in Fig. 3. The variability in the strength of the saturation and residual polarization suggests that the direction of the electric field has a strong influence on ferroelectricity.

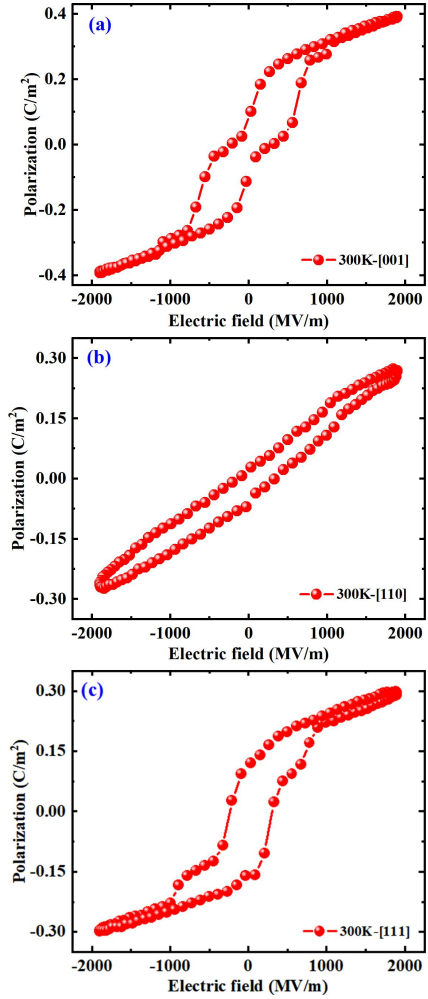

Fig.3 Simulated hysteresis loops of BGO in the directions of (a)[001], (b)[110] and (c)[111]

BGO has a significant anisotropy highlighting its ferroelectricity. The spontaneous polarization is greatest in the [001] direction, indicating that the main polarization occurs in this direction. As can be seen from the Fig.3, the magnitude of the coercive field is approximately the same in all directions, which implies that the difficulty of polarization reversal of ferroelectric domains is also approximately the same.

## 3.3 Flexoelectric effect

To study the flexoelectricity of BGO, BGO nanoribbons were simulated with molecular dynamics, and a three-dimensional cantilever model was designed in which a force of fixed magnitude was applied to a section of the beam. At this point, the flexoelectric effect of the structure can be assessed from the calculated polarization. Typically, the flexoelectric polarization $P_l$ can be written as[10]

$$P_l = \mu_{ijkl} \frac{\partial S_{ij}}{\partial x_k} \quad (i,j,k,l=1,2,3) \qquad (4)$$

where $\mu_{ijkl}$ is the flexoelectric coefficient and $\partial S_{ij}/\partial x_k$ is the strain gradient along the $x_k$ direction. $S_{ij}$ is the strain. In our previous post[11], convert a four-footed label to a two-footed label. The equation becomes

$$P_2 = \mu_{22} \frac{\partial S_{22}}{\partial x_2} \qquad (5)$$

$\mu_{22}$ is the flexoelectric coefficient and $S_{22}$ is the strain along the $x$ direction. $\partial S_{22}/\partial x_2$ is the strain gradient along the y direction.

A cantilever beam model of a BGO capable of stable operation was developed to study the variation of the flexoelectric coefficient of the material at different bending angles. The cantilever beam was further loaded with different magnitudes of electric fields in the bending direction to investigate the specific reasons for the difference in the flexoelectric coefficients between the experiments and the simulations. The cantilever beam was built with dimensions of 2×5×40 and size of 3.138 nm×2.746 nm ×21.532 nm. it was divided into three parts: fixed part, free part, and force part. After several tests, a suitable constant magnitude force is applied to the applied portion to bend the cantilever beam in the y-direction.

The bending angle is adjusted by changing the length of the fixed Fig.4 cells, 8 cells, 12 cells, 15 cells, 19 cells, 22 cells, and 26 cells are fixed to adjust different bending angles in order to investigate the relationship with the flexoelectric coefficient. In a very small range of angles, in which fixed 8 cell,15 cell, and 22 cell were selected as a demonstration, after linear fitting of the data, it was found that the bending angle of the cantilever beam decreases and the deflection coefficient decreases as the length of the fixed part becomes longer, and the deflection coefficient

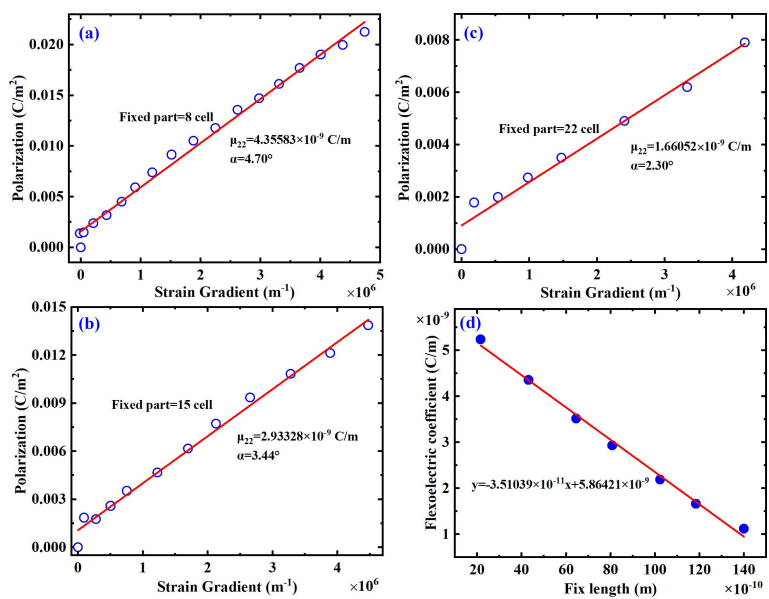

Fig.4 Small angle bending with fixed length of beam. The fixed part was (a) 8 cells (b)15 cells and (c)22 cells respectively, and (d) relationship between the fixed length and flexoelectric coefficient

is related to the length of the fixed part.

To accurately predict the flexoelectric coefficient, a physical process is analyzed essentially[12]

$$\mu = \frac{P}{\alpha} \cdot (L-x) \tag{6}$$

where L is the overall cantilever beam length, P is the polarization and x is the fixed length. The α is the angle corresponding to the arc when bending. When the length of the fixed part is 0, it indicates that the maximum flexoelectric coefficient can be obtained when the cantilever beam is bent as a whole. This linear relationship can be analyzed by Eq.(6). The linear fit was placed in Fig.4(d), and the maximum flexoelectric coefficient of BGO can be predicted to be $5.86\times10^{-9}$ C/m based on the fitted data.

## 3.4 Reasons for the difference between analogue and experimental flexoelectric effect

The flexoelectric equation with electric field action can be written as[13]

$$P = \chi E + \mu \frac{\partial u}{\partial x} \tag{7}$$

During the simulation, there is no need to load a certain electric field, and the formula for the flexoelectric effect is written as

$$P_i = \mu_t \cdot \frac{\partial S}{\partial x_i} \tag{8}$$

In order to analyze the discrepancy between the theoretical and experimental results, we have to make the assumption that bending produces a reversed additional electric field inside the crystal, which could come from quantum effects, thus one gets

$$\mu_t \cdot \frac{\partial S}{\partial x_i} + \varepsilon \cdot E_a = \mu_e \frac{\partial S}{\partial x_i} \tag{9}$$

where $E_a$ is the additional electric field, then it could be obtained as

$$E_a = \frac{(\mu_e - \mu_t)}{\varepsilon} \cdot \frac{\partial S}{\partial x_i} \tag{10}$$

In order to facilitate the estimation of the magnitude of this additional electric field in different materials, it is necessary to keep the strain gradient as a constant value, for example, to make it as 1 m$^{-1}$, from which it is possible to calculate the magnitude of the additional field inside the materials. Since the value of $\mu_t$ is much smaller than $\mu_e$, this leads to the following expression

$$E_a = \frac{\mu_e}{\varepsilon} \cdot \frac{\partial S}{\partial x_i} \tag{11}$$

As can be seen in Table 1, the theoretical values are much smaller than the experimental values. Therefore, the above discussion is reasonable. The magnitude of the additional electric field thus calculated is listed in the last column of the Table 2.

Table 2 Calculated additional electric fields for various materials based on assumptions

| Materials | Source | $\mu_{12}$ (C/m) | Methods | $\varepsilon = \varepsilon_r \cdot \varepsilon_0$ (C·V$^{-1}$·m$^{-1}$) | $E_a$(MV/m) |
|---|---|---|---|---|---|
| BaTiO$_3$ | Gharbi M[15] | 4×10$^{-6}$ | experiment | $\varepsilon_r$=2300 | 1.92×10$^2$ |
|  | Ma W[16] | 5×10$^{-5}$ | experiment |  | 2.48×10$^3$ |
|  | Ma W[17] | 3.8×10$^{-5}$ | experiment |  | 1.8×10$^3$ |
|  | Wenhui Ma[31] | 9×10$^{-6}$ | experiment |  | 4.4×10$^2$ |
| Ba$_{0.7}$Ti$_{0.3}$O$_3$ | Kwon S R[21] | 2.45×10$^{-5}$ | experiment | $\varepsilon_r$=327 | 8.47×10$^3$ |
| BT-8BZT | Huang S J[22] | 2.5×10$^{-5}$ | experiment | $\varepsilon_r$=2700 | 1.05×10$^3$ |
| PMN-PT | Narvaez J[23] | 3.8×10$^{-5}$ | experiment | $\varepsilon_r$=35000 | 1.24×10$^3$ |
| Ba$_{0.67}$Sr$_{0.33}$TiO$_3$ | Ma W H[24] | 1×10$^{-4}$ | experiment | $\varepsilon_r$=13000 | 8.7×10$^2$ |
|  | Ma W H[25] | 8.5×10$^{-6}$ | experiment |  | 7.34×10$^3$ |
| PZT | Ma W H[29] | 1.5×10$^{-6}$ | experiment | $\varepsilon_r$=2100 | 8.02×10$^1$ |
| PMN | Ma W H[23] | 4×10$^{-6}$ | experiment | $\varepsilon_r$=11720 | 3.84×10$^1$ |
|  | W. Ma[30] | 3.4×10$^{-6}$ | experiment |  | 3.27×10$^1$ |
| SrTiO$_3$ | Ponomareva[18] | 3.3×10$^{-9}$ | DFT |  |  |
| BaZrO$_3$ | Hong J W[19] | −2.1×10$^{-10}$ | first principles |  |  |
| PbTiO$_3$ |  | −0.25×10$^{-9}$ |  |  |  |
| PVDF | Hu T[20] | 1.24×10$^{-9}$ | MD |  |  |
| PE |  | −2.63×10$^{-9}$ |  |  |  |
| BaTiO$_3$ | Zhou L[12] | 1.85×10$^{-8}$ | MD |  |  |
|  | Xu T[14] | 1.6×10$^{-9}$ | DFT |  |  |

This result explains the existence of an internal electric field. But the resulting electric field is so strong that it can breakdown the sample through. Therefore, the previous assumption is incorrect. We must find another way out. Besides, the assumption does not give insight into the design of the experiment. According to the theory of mechanics of materials, for a cantilever beam of length $L$ and thickness $h$, if the force applied at the endpoint is $F$, the resulting deflection at the endpoint is $\delta$, then

$$\delta = \frac{F \cdot L^3}{3YI} \quad (12)$$

Where $Y$ is Young's modulus, $I$ is moment of inertia.

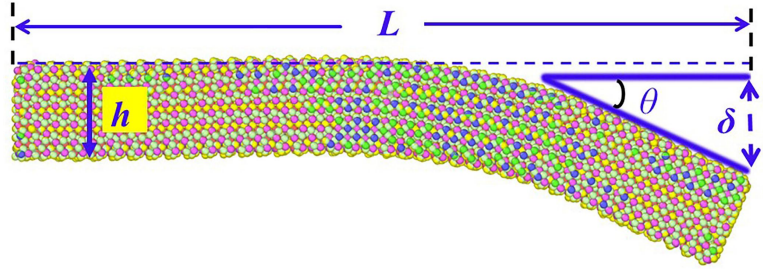

Fig.5 Structural schematic of small bending beam.

As illustrated in Fig.5, the angle generated by the bending of the cantilever beam can be expressed as

$$\theta = \frac{F}{YI} \cdot \frac{L^2}{2} \qquad (13)$$

The strain generated in the longitudinal direction during bending of a cantilever beam is

$$\varepsilon = \frac{3h}{2L^2} \cdot \delta = \frac{\Delta L}{L} \qquad (14)$$

In the above equation, $\Delta L$ is the elongation length of the cantilever beam. This gives the radius of curvature of the cantilever beam bending as

$$R = \frac{\Delta L}{\theta} = \frac{3h \cdot YI \cdot \delta}{FL^3} \qquad (15)$$

According to equation (4), the following result can be obtained

$$P \propto \frac{\mu}{h} \qquad (16)$$

This means that the thickness $h$ is an important influence for the flexoelectric effect. Here, we assume that four factors have an effect on the polarization, and the function is expressed as $P(h,EI,F,\mu)$. Following the theory of dimensional analysis, the dimensional matrix was created and listed in Table 3.

Table 3 Dimensional matrix for the analysis of $P(h,EI,F,\mu)$

| Symbol | h | YI | F | $\mu$ | P |
|---|---|---|---|---|---|
| Length L | 1 | 3 | 1 | -1 | -2 |
| Mass M | 0 | 1 | 1 | 0 | 0 |
| Time T | 0 | -2 | -2 | 1 | 1 |
| Electric current I | 0 | 0 | 0 | 1 | 1 |

As can be seen from equations (12) and (13), it is more convenient to consider $YI$ as an overall factor, as shown in Table 3. The indices here are given by the SI Units. According to Buckingham's $\pi$(Pi) Theorem, there are two conclusions can been obtained

$$P = \pi_1 \cdot h^{-1} \cdot \mu \qquad (17)$$

$$F = \pi_2 \cdot h^{-2} \cdot YI \qquad (18)$$

here $\pi_1$ and $\pi_2$ are dimensionless physical quantities. Equation (17) is consistent with the conclusion of equation (16). Based on the two equations above, the following function can be obtained

$$\frac{P \cdot h}{\mu} = \varphi\left(\frac{F \cdot h^2}{YI}\right) \tag{19}$$

For a cantilever beam, the load force $F$ is

$$F = \frac{3YI \cdot \delta}{L^3} \tag{20}$$

Substituting the expression of $F$ into equation (19) yields

$$\frac{P \cdot h}{\mu} = \varphi\left(\frac{3\delta \cdot h^2}{L^3}\right) \tag{21}$$

Here $\varphi$ can be an arbitrary function that needs to be determined experimentally. In the current simulation, the above equation is expressed as

$$\frac{P_i \cdot h}{\mu} = \varphi\left(\frac{3\delta_i \cdot h^2}{L^3}\right) \tag{22}$$

The footer $i$ represents the number of steps in the simulation. A graph based on the equation (22) is shown in Figure 6.

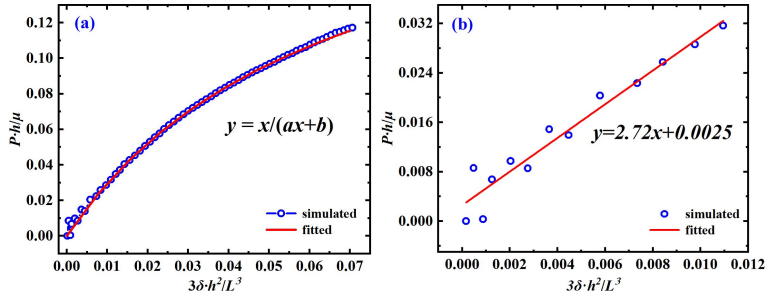

Fig.6 Small angle bending with fixed length of beam.

Figure 6(a) demonstrates that the flexoelectric effect exhibits a nonlinear relationship at larger bending states. We find a simpler function to explore this nonlinear effect as follows

$$y = \frac{x}{ax + b} \tag{23}$$

where $0 \leqslant x < 1$, $a$ and $b$ are the fitting parameters. When $x$ is small, e.g. $x<0.03$, the linear relationship is more pronounced. In particular, the following relationship holds significantly when $x<0.01$

$$\frac{Ph}{\mu} = k \cdot \frac{3\delta \cdot h^2}{L^3} \tag{24}$$

Here $k$ is a scale factor with no units, which is assumed that it is merely related to the material itself and not to the size. In this work, it is 2.72(in Fig.6(b)). For ferroelectric ceramic samples, the degree of bending is experimentally very small, which agrees

well with the current hypothesis. From this it is possible to obtain

$$P = 3k\mu \cdot \frac{h}{L^3} \cdot \delta \tag{25}$$

$$\mu = \frac{P}{3k\delta} \cdot \frac{L^3}{h} \tag{26}$$

Equation (25) shows that the polarization is positively correlated with the deflection, which is consistent with the facts. Furthermore, considering equation (21) from the point of view of dimensionless analysis, since the variables $\delta$ and $h$ do not appear in the table of dimensional matrix and they have the same unit, The following two expressions are equivalent to each other

$$\frac{3\delta \cdot h^2}{L^3} \Leftrightarrow \frac{3\delta \cdot h^2}{L^3} \cdot \left(\frac{\delta}{L}\right)^m \tag{27}$$

here $m$ is a natural number. It can be regarded as a scaling factor, which marks differences in bending models, and its value can be determined experimentally. when the linear relationship holds, the following equation appears

$$\frac{Ph}{\mu} = k \cdot \frac{3\delta \cdot h^2}{L^3} \cdot \left(\frac{\delta}{L}\right)^m \tag{28}$$

This leads to the following equation

$$P = 3k \cdot \mu \cdot \frac{h}{L^{m+3}} \cdot \delta^{m+1} \tag{29}$$

$$\frac{\partial S}{\partial x_i} = 3kh \cdot \frac{\delta^{m+1}}{L^{m+3}} \tag{30}$$

The equations above have rich physical connotations. These equations obtained from the current analysis is universal and can be used in a wide range of bending models. Then, it is useful to define the following representation as the geometric structure factor

$$\gamma = \frac{h}{L^{m+2}} \tag{31}$$

now, equation (29) become

$$P = 3k\gamma \cdot \mu \cdot \frac{\delta^{m+1}}{L} \tag{32}$$

finally, we can use the above equations to explain the difficulties and problems in the current reported literature.

At very small bending, the last term of the above equation scales with the strain gradient. First discuss the case m=0. In simulations, if the thickness of the film is 5 nm and the length is 500 nm, the value of $\gamma$ is $2\times10^{-5}$. But for an experiment, if the thickness of the film is 2 μm and the length is 2 cm, $\gamma$ is obtained as $5\times10^{-9}$. Then let m=1, we get $\gamma$ is $4\times10^{-8}$ and $2.5\times10^{-13}$ respectively. Since the geometric structure factor used in the experiment is very smaller, the flexoelectric coefficients obtained are then much larger. This explains why the experimentally measured flexoelectric coefficients are several orders of magnitude larger than simulated and theoretical. For the cantilever beam, the strain gradient can been expressed by

$$\frac{\partial S}{\partial y} = R\left(\frac{\theta}{L}\right)^2 \tag{33}$$

combining the previously listed equations gives

$$\frac{\partial S}{\partial x_i} = \frac{9h\delta^2}{4L^4} \tag{34}$$

comparison with equation (30) yields $m=1$ and $k=3/4$. This also proves that the results of the previous dimensionless analysis are correct. Based on this theory, new flexural electric coefficient can be calculated.

It can be seen here that the geometrical structure factor plays a very important and decisive role in the flexoelectric effect, but has never been considered in previous studies. Based on the results above, the following two insights can be obtained. First, considering the complexity of strain gradient calculation, the strain gradient can be replaced by the deflection. Such a substitution can simplify all the flexoelectric problems, both theoretical and experimental. On the other hand, the flexoelectric effect has a significant size effect, which is mainly manifested in the geometrical structure factor. This points the way to the experimental discovery of superior flexoelectric materials and the design of flexoelectric devices.

## 4. Conclusions

In this paper, the possible causes of the experimental flexoelectric coefficient and the simulated flexural electric coefficient are explored using the BGO as an example, bridging the gap between experiment and simulation.